\documentclass[aps,pre,preprint]{revtex4-1}

\usepackage{graphicx}
\usepackage{subfig}
\usepackage{subfloat}
\usepackage{dcolumn}
\usepackage{bm}

\usepackage[utf8]{inputenc}
\usepackage[T1]{fontenc}
\usepackage{mathptmx}
\usepackage{amsmath}
\usepackage{etoolbox}
\usepackage{hyperref}
\usepackage{xcolor}

\draft 

\begin{document}


\title{Using test particle sum rules to construct accurate functionals in classical Density Functional Theory} 


\author{Melih G\"ul}
\email[]{melih.guel@uni-tuebingen.de}
\author{Roland Roth}
\email[]{roland.roth@uni-tuebingen.de}
\affiliation{ Institute for Theoretical Physics, University of Tübingen, Auf der Morgenstelle 14, 72076 Tübingen, Germany}
\author{Robert Evans}
\email[]{bob.evans@bristol.ac.uk}
\affiliation{H. H. Wills Physics Laboratory, University of Bristol, Bristol BS8 ITL, United Kingdom}

\date{\today}

\begin{abstract}
Fundamental Measure Theory (FMT) is a successful and versatile approach for describing the properties of the hard-sphere fluid and hard-sphere mixtures within the framework of classical density functional theory (DFT). Lutsko [Phys. Rev. E {\bf 102}, 062137 (2020)] introduced a version of FMT containing two free parameters, to be fixed by additional physical constraints. Whereas Lutsko focused on the stability of crystalline phases, we introduce and employ two statistical mechanical sum rules pertinent for the fluid phase, that are not automatically satisfied by FMT. By minimizing the relative deviation between different routes to calculate the excess chemical potential and the isothermal compressibility we determine the two free parameters of the theory. Our results indicate that requiring consistency with these sum rules can improve the quality of predictions of FMT for properties of the hard-sphere fluid phase. We suggest that employing these (test particle) sum rules, which apply for any interparticle pair-potential, might provide a means of testing the performance and accuracy of general DFT approximations. 
\end{abstract}


\maketitle 
\newpage
\section{Introduction}

Classical density functional theory (cDFT) \cite{cDFT-functional,Hansen-McDonald} provides a powerful framework for investigating the structure and thermodynamic properties of inhomogeneous many-particle systems subject to an external potential and in thermal equilibrium with a heat reservoir. The key object in cDFT is the excess Helmholtz free energy functional ${\cal F}_{ex}[\rho]$, which contains all the information about interactions between particles. However, constructing this object remains a non-trivial task that requires approximations. A breakthrough in cDFT was the introduction of fundamental measure theory (FMT) for hard-sphere mixtures \cite{Free-energy-model-of-hard-spheres} which built upon the form of the only exact cDFT, i.e. that for a one-dimensional system of hard rod mixtures \cite{Percus76,1d-hard-rod}. The formalism makes use of so-called weighted densities that are defined as convolutions of the local densities and weight functions which represent geometrical features of the constituent particles.

This approach inspired further developments of FMT for hard-sphere mixtures in three dimensions.  Incorporating more accurate equation of states led to what are regarded as state of the art, namely the White-Bear functionals \cite{FMT-for-hard-sphere-mixtures, Structures-of-hard-sphere-fluids-from-mod.-FMT, WB-mark-II}. Dimensional crossover \cite{FMT-and-dimensional-crossover} constitutes another attempt to construct an excess free energy functional for hard-sphere systems by requiring that a general ansatz for ${\cal F}_{ex}[\rho]$, say in three dimensions, should yield an accurate lower-dimensional limit when restricted appropriately.Introducing dimensional crossover was an important step in accounting for the freezing transition of hard spheres, which cannot be described by the original Rosenfeld functional.

Lutsko \cite{esMFT-Lutsko} revisited the concept of dimensional crossover focusing on the stability of the functional, especially in the context of crystalline phases. The construction relies on the systematic approach introduced by Rosenfeld and Tarazona \cite{Tarazona-Rosenfeld-1997}. Lutsko showed that demanding explicit stability of the functional, i.e. that the free energy density is bounded from below, inevitably leads to less accurate results at low densities. Restricting the analysis to weights already employed in FMT, a more general excess functional ${\cal F}_{ex}[\rho]$ was derived, which we refer to as the Lutsko functional. This builds upon the Rosenfeld functional and is determined up to two free parameters, named $A$ and $B$. Explicit stability of the corresponding functional requires $A$ and $B$ to be positive or zero. Lutsko considered the choice $A=1.0$ and $B=0.0$ and showed that the accuracy of the corresponding functional is comparable to that of White-Bear \cite{esMFT-Lutsko}. 

In this paper we investigate some the properties of the Lutsko functional for the hard-sphere fluid phase in the light of statistical mechanical sum rules. These play a crucial role when it comes to examining the consistency of a given (approximate) functional. Generally sum rules relate microscopic (structural) properties of a system to thermodynamic ones. For instance, the contact sum rule connects the one-body density profile $\rho(\mathbf{r})$ at contact with a planar hard wall to the bulk-pressure $P$ of the system whilst the Gibbs adsorption sum rule relates the excess adsorption at a wall, obtained from the density profile, to the wall-fluid surface tension. Both sum rules are automatically satisfied by FMT and hence are routinely employed to test the accuracy of numerical implementation of the DFT minimisation.

Here we introduce two additional sum rules for testing the consistency and accuracy of cDFT: (i) a sum rule for the excess chemical potential $\mu_{ex}$ , that can be calculated from either the bulk excess free energy density or from the inhomogeneous density distribution $\rho(r)$ of a fluid in the external potential due to a fixed fluid particle, i.e. in test-particle geometry, and (ii) for the isothermal compressibility $\chi_T$, which can be calculated from the bulk equation of state or by a suitable integral over the inhomogeneous density distribution $\rho(r)$ in the same test-particle geometry. The sum rules are valid for an arbitrary pair potential $\phi(r)$. Of course, if one knew the exact free energy functional properties calculated by the different routes would be identical. This is not the case for FMT approximations, i.e. using different routes to calculate either the excess chemical potential or the isothermal compressibility is not guaranteed to yield the same result for the property under investigation. Indeed we expect to find differences. This observation allows us to employ these two sum rules as a means to  determine the parameters $A$ and $B$ of the Lustko functional. We focus on the fluid phase. 
  
Our paper is organised as follows. We introduce the theoretical framework in Sec.~\ref{sec:theory}, including in IIA some background to cDFT and a description of the pertinent sum rules. FMT, the Lutsko functional and its implementation in test particle geometry are described in the subsequent sub sections. Sec.~\ref{NumMeth} describes briefly how our numerical calculations are performed. In Sec.~\ref{results} we present the results of our study that is aimed at obtaining optimal values for $A$ and $B$ in the Lutsko functional by minimizing deviations between the different (test particle) predictions for (i) $\mu_{ex}$ and (ii) $\chi_T$ across the full range of fluid densities. By comparing results for the equation of state and for density profiles at a spherical test particle and at a planar hard wall with simulation and existing FMT approaches we assess the overall performance of the Lutsko functional in the fluid phase. We conclude in Sec.~\ref{sec:conclusion} with a summary and an outlook.

\section{Theory} \label{sec:theory}
\subsection{DFT and importance of Sum Rules}\label{DFT-sum-rules}
In classical density functional theory (DFT) the grand-canonical functional, of the one-body density distribution $\rho(\mathbf{r})$, is
\begin{equation}\label{Omega-Functional}
    \Omega[\rho] = {\cal F}_{id}[\rho] + {\cal F}_{ex}[\rho] + \int d\mathbf{r}\,\rho(\mathbf{r})\left(V_{ext}(\mathbf{r})-\mu\right).
\end{equation}
This functional is minimized for the equilibrium density distribution $\rho_{eq}(\mathbf{r})$ \cite{cDFT-functional}, i.e.
\begin{equation}\label{Omega-minimize}
    \frac{\delta\Omega[\rho(\mathbf{r})]}{\delta\rho(\mathbf{r})}\bigg|_{\rho(\mathbf{r})=\rho_{eq}(\mathbf{r})} = 0,
\end{equation}
where we note ${\cal F}_{id}[\rho]$ is the ideal and ${\cal F}_{ex}[\rho]$ is the excess Helmholtz free energy functional.The latter describes the interactions between the particles. $V_{ext}(\mathbf{r})$ is the external potential and $\mu$ is the chemical potential. $\Omega[\rho_{eq}]=\Omega$ is the grand potential of the inhomogeneous fluid.

The exact ${\cal F}_{ex}[\rho]$ is, of course, unknown and in practice one must find reliable approximations for a given choice of (model) fluid, see e.g. the reviews \cite{DFT-of-Non-uniform-fluids} and \cite{Recent-Develops-Lutsko}. Finding suitable approximations is often guided by physical intuition. However, the consistency of any approximation can be ascertained by calculating thermodynamic or structural quantities by different routes and checking that these yield the same results.
Two standard sum rules are routinely examined within DFT. The first, for the spherical geometry considered in this paper, reads
\begin{equation}\label{adsorption-sum-rule}
    \Gamma(R_s,\mu)=4\pi\int_{R_s}^\infty dr\,r^2(\rho(r)-\rho_b)=-\left(\frac{\partial\Omega^s}{\partial\mu}\right)_T,
\end{equation}
where the Gibbs excess adsorption $\Gamma$ depends on the radius $R_s$ of a hard spherical 'solute', $\rho_b$ is the density of the fluid in the bulk, at given ($\mu,T$), far from the solute and $\Omega^s$ is the surface excess grand potential in the accessible volume. Calculating the density profile by minimizing the (approximate) $\Omega[\rho]$ and integrating over the accessible volume should yield the same value of $\Gamma$ as is obtained from the derivative w.r.t. $\mu$ of the (approximate) $\Omega$. Most, but not all, DFT approximations will automatically  satisfy
Eq.\eqref{adsorption-sum-rule}.
The second sum rule is of a very different character. For a hard spherical 'solute' of radius $R_s$ the contact value of the density profile is given by \cite{Henderson83,Bryk03}
\begin{equation}\label{contact-sum-rule}
    k_B T\rho(R_s^+)=P+\frac{1}{4\pi R_s^2}\left(\frac{\partial\Omega^s}{\partial R_s}\right)_{T,\mu},
\end{equation}
where $P$ is the pressure in the bulk fluid at given ($\mu,T)$. In planar geometry ($R_s=\infty$) Eq.\eqref{contact-sum-rule} reduces to the well-known result $k_B T\rho(0^+)=P$, for the contact value $\rho(0^+)$ of the density profile at a planar hard wall.
Most (sensible) non-local DFT approximations will satisfy the contact sum rule automatically. Thus, comparing each side of Eq.\eqref{contact-sum-rule} does not provide a measure of how reliable a certain DFT approximation might be. Rather such comparisons Eq.\eqref{adsorption-sum-rule} and Eq.\eqref{contact-sum-rule} merely provide a valuable check on the accuracy of the numerics used in the DFT calculation.

Here we identify two other sum rules, valid for any pair potential, that do test the consistency and the accuracy of DFT approximations. Both are based on two classic papers by Percus \cite{Percus-1962,frisch1964equilibrium}; see also\cite{Hansen-McDonald}. Percus proved that the grand partition function for a fluid with an additional test particle exerting an 'external' potential $v(r)\equiv\phi(r)$, the pairwise interparticle potential, is given by
\begin{equation}\label{grand-partition-function}
    \Xi(\phi)=\frac{\rho_b}{z}\,\Xi(\phi=0),
\end{equation}
where the activity $z=e^{\beta\mu}\Lambda^{-3}$, with $\beta=(k_B T)^{-1}$ and $\Lambda$ the thermal de Broglie wavelength. $\Xi(\phi=0)$ refers to the partition function in the absence of the additional particle.
It follows that the excess grand potential for this test particle situation is
\begin{align}\label{excess-grand-pot}
    \Omega^s_{\text{test}}&=-k_B T\left(\log\Xi(\phi)-\log\Xi(0)\right)\\\nonumber
    &=\mu-k_B T\log(\rho_b\Lambda^3)\equiv\mu_{ex},
\end{align}
where $\mu_{ex}$ is the excess over ideal chemical potential. Eq.\eqref{excess-grand-pot} is particularly important within the context of DFT.
Suppose one fixes a test particle at the origin, exerting a potential $v(r)$, then employs an (approximate) DFT, determines the equilibrium profile $\rho(r;\phi)$ and measures the corresponding (approximate) grand potential. Eq.\eqref{excess-grand-pot} then yields a result for $\mu_{ex}$ for the uniform (bulk) fluid. There is no reason to suppose that this test particle route should yield a value identical to that obtained by calculating $\mu$ from the original free energy functional, evaluated with a constant bulk density $\rho_b$, i.e. from $\mu(\rho_b)=(\partial f/\partial\rho_b)_T$, where $f(\rho_b)$ is the free energy density of the bulk. The magnitude of the difference between the two routes to $\mu_{ex}$ should reflect the consistency of the approximate DFT.
\begin{figure}
	\includegraphics[scale=0.45]{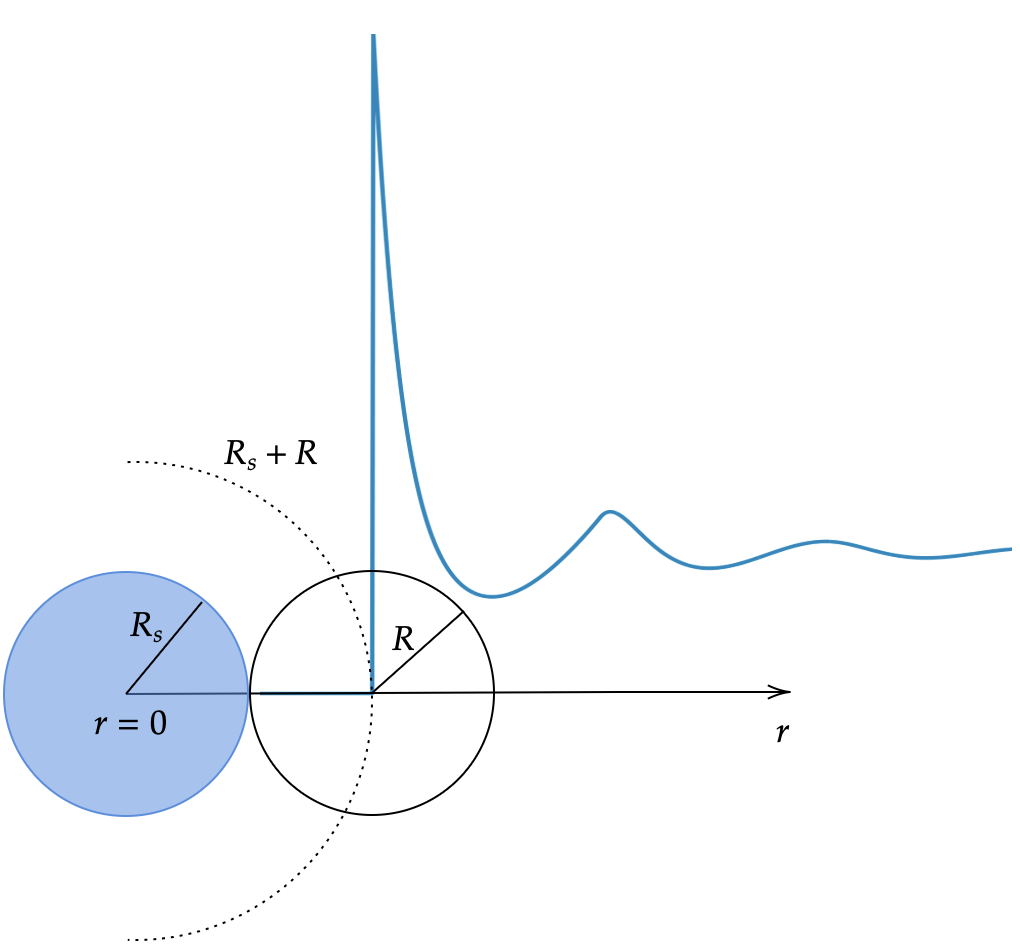}
	\caption{\label{fig:test-particle} Schematic of test-particle geometry for hard spherical particle. A test particle of radius $R_s$ (blue) placed at the origin $r=0$ interacts with neighboring particles of radius $R$ leading to an exclusion sphere of radius $R_s+R$. In the test-particle limit, $R_s=R$, the one-body density profile (blue curve) is proportional to the RDF $g(r)$ of the bulk fluid; see Eq.(7).}
\end{figure}

The second important result that Percus\cite{Percus-1962} proved is that the one-body density $\rho(r;\phi)$ is proportional to the radial distribution function (pair-correlation function) $g(r)$ of the uniform fluid, i.e.
\begin{equation}\label{radial-dist-func}
    \rho(r;\phi)=\rho_b\,g(r).
\end{equation}
Recall that the bulk density is fixed by ($\mu,T$). Eq.\eqref{radial-dist-func} is what one expects intuitively, and it corresponds to how one often explains the form of $g(r)$ ; think of fixing a coin at the origin and arranging the (identical) coins around this.
However, the content is non-trivial as it relates a particular one-body distribution of an inhomogeneous fluid to the two-body (density-density) correlation function of the homogeneous (bulk) fluid.
It follows from Eq.\eqref{radial-dist-func} that
\begin{equation}\label{structure-factor}
    \int d\mathbf{r}\,(\rho(r;\phi)-\rho_b)=\rho_b\int d\mathbf{r}\,(g(r)-1)=-1+S(k=0),
\end{equation}
where the integral is over all space and $S(k)$ is the usual static structure factor, e.g. see \cite{Hansen-McDonald}.
In the limit of vanishing wavenumber $S(k=0)$ is proportional to the bulk isothermal compressibility $\kappa_T$
\begin{equation}\label{iso-compress}
    S(k=0)=\rho_b k_B T\kappa_T = \chi_T,
\end{equation}
where  $\chi_T$  is defined as the reduced compressibility. Of course, in an exact treatment this test particle route to the compressibility is guaranteed to be exact. This is not the case for an approximate DFT. The density profile $\rho(r;\phi)$, obtained by minimizing the approximate functional, will yield via integration a value for $S(k=0)$, and hence $\kappa_T$. There is no reason to expect that this value of $\kappa_T$ will be identical to that obtained directly from the bulk free energy density, i.e. using $\rho_b^2\kappa_T=(\partial\mu/\partial\rho_b)^{-1}$.

Once again, differences between test-particle results and those obtained from the bulk free energy density should provide information about the consistency and, possibly, the accuracy of a given DFT approximation.
The test-particle geometry is illustrated in Fig.\ref{fig:test-particle} for the case of a hard spherical particles; it corresponds to the limit where the 'solute' radius $R_s=R$ ,the radius of the 'solvent' particles.
Generally, the test-particle limit corresponds to making the solute particle identical to the solvent particles.
\subsection{Fundamental Measure Theory for Hard Spheres}
Fundamental measure theory (FMT) provides a powerful framework for understanding the properties of inhomogeneous fluids with a hard-core repulsion \cite{FMT-for-hard-sphere-mixtures}. The excess free energy functional ${\cal F}_{ex}[\rho]$ for such systems can be constructed using a set of geometrical measures describing the size and shape of each species. 

We focus here on hard-sphere (HS) systems for which Rosenfeld made the following ansatz for the excess free energy functional \cite{FMT-for-hard-sphere-mixtures, Free-energy-model-of-hard-spheres} 
\begin{equation}
    \beta{\cal F}_{ex}[\rho] = \int d\mathbf{r}\,\Phi_{ex}(\{n_\alpha\}),
\end{equation}
where $\Phi_{ex}$ is the free energy density, which is a function of the set of weighted densities $n_\alpha$. For a 
 one-component system these are defined as 
convolutions of the corresponding weight functions $w_{\alpha}(\mathbf{r})$ with the local density distribution $\rho(\mathbf{r})$ :
\begin{equation}
    n_\alpha(\mathbf{r}) = \int d\mathbf{r}'\,\rho(\mathbf{r})w_\alpha(\mathbf{r}-\mathbf{r}').
\end{equation}
For a hard-sphere fluid with radius $R$ the associated weight functions read
\begin{align}
    w_3(\mathbf{r}) = \Theta(R-|\mathbf{r}|),&\quad w_2(\mathbf{r}) = \delta(R-|\mathbf{r}|)\\\nonumber
    w_1(\mathbf{r}) = \frac{w_2(\mathbf{r})}{4\pi R},&\quad w_0(\mathbf{r}) = \frac{w_2(\mathbf{r})}{4\pi R^2}\\\nonumber
    \mathbf{w}_2(\mathbf{r}) = \delta(R-|\mathbf{r}|)\mathbf{e}_r&,\quad \mathbf{w}_1(\mathbf{r}) = \frac{\mathbf{w}_2(\mathbf{r})}{4\pi R},
\end{align}
where the first four weights are scalar and the subsequent two are vectors. We focus here on the one-component fluid ; generalization to mixtures is straightforward.

The original Rosenfeld functional (RF) \cite{Hard-sphere-crystal,Free-energy-model-of-hard-spheres} was derived using a combination of dimensional analysis, the exact low-density expansion and scaled-particle theory \cite{SPT-of-fluid-mixtures}. The excess free energy density takes the form
\begin{equation}\label{RF}
	\Phi_{ex}^{\text{RF}} = \Phi_1+\Phi_2+\Phi_3^{\text{RF}},
\end{equation}
where
\begin{align}\label{RF-functional}
	\Phi_1 &= -n_0\log(1-n_3)\\\nonumber
	\Phi_2 &=\frac{n_1 n_2-\mathbf{n}_1\cdot\mathbf{n}_2}{1-n_3}\\\nonumber
	\Phi_3^{\text{RF}} &= \frac{n_2^3-3n_2\mathbf{n}_2\cdot\mathbf{n}_2}{24\pi(1-n_3)^2}.
\end{align}
Rosenfeld's HS functional is constructed so that taking two functional derivatives w.r.t. density  yields the Percus-Yevick (PY) pair-direct correlation functions $c^{(2)}_{ij}(r)$ in the bulk mixture. It follows that the bulk equation of state (EoS) is given by the PY compressibility EoS, which for the one-component case is \cite{PY-EoS}
\begin{equation}\label{PY-EoS}
	\frac{\beta P^{\text{PY}}}{\rho} = \frac{1+\eta+\eta^2}{(1-\eta)^3},
\end{equation}
Here $\eta$ is the packing fraction $\eta=\rho v$, where $v$ is the HS particle volume. It is well-known that this EoS is in good agreement with simulations for hard-sphere fluids at low and medium packing fractions but overestimates the pressure when approaching the freezing transition. 

The Mansoori-Carnahan-Starling-Leland (MCSL) EoS \cite{Carnahan69,MCSL-EoS} is more accurate than PY for HS mixtures and was used as an input into the White-Bear (WB) functional \cite{FMT-Revisited-and-WB, Structures-of-hard-sphere-fluids-from-mod.-FMT}. Incorporating an even more accurate EoS led to the mark II version of the WB-functional \cite{WB-mark-II}.
Generally, the WB functional performs better than the original Rosenfeld functional for a one-component HS fluid; the improvement can be traced to the fact that the bulk EoS corresponds to the accurate Carnahan-Starling (CS) approximation rather than PY. For example, this ensures that the contact value of the density profile at a hard wall is given more accurately by WB than by RF, see \cite{FMT-for-hard-sphere-mixtures}.

The RF and WB functionals automatically satisfy the Gibbs adsorption and contact sum rules. It is important to note that both functionals contain no free parameters. When we examine their performance, regarding the degree to which the two test particle sum rules are satisfied,  there is no prescription that will allow us to use consistency to attempt to improve the functional.
\subsection{The Lutsko HS Functional} \label{sec:LK-Functional}
It was recognized by Rosenfeld in his original paper \cite{Free-energy-model-of-hard-spheres} that the RF functional fails to account for the HS freezing transition. Crystallization can be viewed as a special case of strong confinement : particles remain localized near their lattice sites.  By considering particular classes of confinement FMT was extended so that crystallization could be incorporated. For a higher dimensional system exact results for lower dimensions (D) result can be incorporated and was termed "dimensional crossover" \cite{FMT-and-dimensional-crossover}. Recall that for the 1D hard-rod system and  the 0D-cavity system, which can at most hold one particle, the functional is known exactly. The reviews by Roth \cite{FMT-for-hard-sphere-mixtures} and Tarazona et.al. \cite{Tarazona2008} describe developments. Tarazona and Rosenfeld \cite{Tarazona-Rosenfeld-1997} provided important new insight into the geometrical constraints required by different cavity shapes.

Recently Lutsko re-cast  insightfully the ideas of Tarazona and Rosenfeld  and proposed the following functional for a one-component  3D hard-sphere system \cite{esMFT-Lutsko}
\begin{equation}\label{Lutsko-Functional}
    \Phi^{\text{LK}}_{ex} = \Phi_1+\Phi_2+\Phi^{\text{LK}}_3,
\end{equation}
where $\Phi_1$ and $\Phi_2$ are identical to those of Rosenfeld's functional Eq.\eqref{RF-functional}. The third term has a different structure with two additional parameters $A$ and $B$
\begin{widetext}
\begin{equation} \label{eq:phi3_LK}
    \Phi_{3}^{\text{LK}} = \frac{(A+B)n_2^3-3 A~ n_2 \mathbf{n}_2\cdot\mathbf{n}_2+3 A~ \mathbf{n}_2~\mathbf{T}~\mathbf{n}_2-3 B~n_2\text{Tr}\,\mathbf{T}^2+(2 B-A)\text{Tr}\,\mathbf{T}^3}{24\pi(1-n_3)^2},
\end{equation}
\end{widetext}
where $\text{Tr}$ denotes the trace and the tensorial weighted density is
\begin{equation}
    \mathbf{T}(\mathbf{r}) = \int d\mathbf{r}'\,\rho(\mathbf{r}) w_T(\mathbf{r}-\mathbf{r}').
\end{equation}
with the tensorial weight defined as
\begin{align}\label{tensor-weight}
    w_T(\mathbf{r}) = \delta(&R-|\mathbf{r}|)\mathbf{e}_r\otimes \mathbf{e}_r.
\end{align}
Here, $\mathbf{e}_r = \mathbf{r}/r$ is the radial unit vector and $\otimes$ denotes the dyadic product.
Note that Eq.~(\ref{eq:phi3_LK}) is a generalisation of the corresponding expression for $\Phi_3$ of the tensor version \cite{Tarazona-Rosenfeld-1997} of the Rosenfeld functional \cite{Free-energy-model-of-hard-spheres}, which is recovered for $A=-B=3/2$.

The parameters $A$ and $B$  must be determined by imposing physical constraints. Lutsko was concerned with functionals that could lead to stable crystals and focused on global stability. He showed that for $A, B\geq 0$ the functional is automatically stable, i.e. the free energy is bounded from below \cite{cDFT-Lutsko-on-stability} and termed this class of functional "explicitly stable" or  esFMT$(A,B)$ \cite{esMFT-Lutsko}.

The bulk EoS  from the Lutsko HS functional reads
\begin{equation}\label{LK-EoS}
    \frac{\beta P^{\text{LK}}(A,B)}{\rho} = \frac{\beta P^{\text{PY}}}{\rho} + \frac{\eta^2}{3(1-\eta)^3}\left(8A+2B-9\right).
\end{equation}
Clearly the PY EoS is recovered for $8A+2B-9 =0$ , defining the PY-line in the parameter plane $(A, B)$. In the following we shall abbreviate Lutsko's functional as LK$(A, B)$.

\subsection{Implementing the Test-Particle Sum Rules within DFT}\label{sum-rules}

We examine the sum rules for the excess-chemical potential $\mu_{ex}$ and reduced isothermal compressibility $\chi_T$ set out in Sec.\ref{DFT-sum-rules} and enquire whether these might be used to determine $A$ and $B$ , the parameters introduced by Lutsko. Recall that both thermodynamic quantities can be obtained directly from  the bulk free energy density, which we term the bulk thermodynamic route, or by  performing a numerical computation of the density profile within DFT. If the associated functional were exact the two routes would yield the same value for $\mu_{ex}$ and for $\chi_T$. As we shall  observe,  for HS functionals such as RF or WB , there are deviations that become severe for higher densities. 
Therefore, we search for parameters $A$ and $B$ such that the associated LK functional satisfies best these sum rules , i.e. values for which the deviations are minimal.

The numerical DFT computation is based on the test-particle procedure  in which we hold a particle of radius $R_s$ in a fixed position $r=0$ surrounded by other particles of radius $R$, see Fig.\ref{fig:test-particle}. Spherical symmetry dictates that the underlying calculation is one-dimensional in the radial distance $r$. 

The external potential $V_{ext}(r)$ in Eq.\eqref{Omega-Functional} is
\begin{equation}
    V_{ext}(r) = \begin{cases}
        \infty,\quad r< R_s+R,\\
        0,\quad \text{otherwise}.
    \end{cases}
\end{equation}
We minimize the grand potential functional in the test particle limit  $R_s=R$. From the resulting equilibrium density profile $\rho(r)\equiv \rho(r;\phi)$ we obtain the radial distribution function (RDF) $g(r)=\rho(r)/\rho_b$ and and the corresponding grand potential , Eq.\eqref{Omega-minimize} and below. Specifically we calculate $\mu_{ex}$ using Eq.(6) so that
\begin{equation}
\mu_{ex}^{\text{DFT}}=\Omega_1-\Omega_0,
\end{equation}
where $\Omega_0=-PV$ is the grand potential before insertion and $\Omega_1=\Omega[\rho(r)]$ that after insertion.
$\chi_T$ is obtained using Eqs.( 8 and 9). As these quantities are  obtained from $\rho(r)$ employing DFT we denote these as $\mu_{ex}^{\text{DFT}}$ and $\chi_T^{\text{DFT}}$. 

The excess chemical potential $\mu_{ex}$ from the bulk thermodynamic route is given by
\begin{equation}
	\beta\mu_{ex}^{\text{Bulk}} = \frac{\partial\Phi_{ex}}{\partial\rho}\bigg|_{\rho=\rho_b},
\end{equation}
evaluated at bulk density $\rho_b$. Inserting the free energy density of Lutsko Eq.\eqref{Lutsko-Functional} yields 
\begin{equation}\label{mu-bulk}
    \beta\mu^{\text{Bulk}}_{ex}(A,B) = \beta\mu_{ex}^{\text{RF}}+\frac{\eta^2(3-\eta)}{6(1-\eta)^3}(8A+2B-9),
\end{equation}
where
\begin{equation}
	\beta\mu_{ex}^{\text{RF}} = -\log(1-\eta)+\frac{\eta(14-13\eta+5\eta^2)}{2(1-\eta)^3}
\end{equation} 
is the excess chemical potential of RF, identical to that from the PY compressibility route.

The reduced isothermal compressibility $\chi_T$ is obtained from the EoS of Lutsko Eq.\eqref{LK-EoS}
\begin{equation}\label{xt-bulk}
    \chi^{\text{Bulk}}_T(A,B) = \chi_T^{\text{RF}}\frac{(1+2\eta)^2}{(1+2\eta)^2+(8A+2B-9)\eta^2},
\end{equation}
where
\begin{equation}
	\chi_T^{\text{RF}} = \frac{(1-\eta)^4}{(1+2\eta)^2}
\end{equation}
is the isothermal compressibility of RF identical to the PY result. As was the case for $P^{\text{LK}}$ we see that if $8A+2B-9=0$ both $\mu_{ex}^{\text{Bulk}}$ and $\chi_T^{\text{Bulk}}$ reduce to the RF expressions. Thus, the magnitude of $8A+2B-9$ 
 determines how much the corresponding thermodynamic quantity deviates from its PY value. 

The quantities we chose for minimization are defined as
\begin{equation}\label{relative-devs}
    \delta_\mu = \frac{\mu^{\text{DFT}}_{ex}-\mu^{\text{Bulk}}_{ex}}{\mu^{\text{Bulk}}_{ex}},\quad
    \delta_\chi = \frac{\chi_T^{\text{DFT}}-\chi_T^{\text{Bulk}}}{\chi_T^{\text{Bulk}}}
\end{equation}
We obtain optimal values for the parameters $A$ and $B$  with the minimization prescription: $\underset{\eta,\,A,\,B}{\text{min}}\,M(\delta_\mu,\delta_\chi)$ where $M$ is a suitable function of $\delta_\mu$ and $\delta_\chi$ for the optimization process.

\section{Numerical Methods}\label{NumMeth}
As mentioned in Sec.\ref{sum-rules} we perform DFT in the test-particle geometry where calculations such as convolutions are performed in spherical coordinates. The weighted densities $n_\alpha(\textbf{r})$ and the one-body direct correlation function $c^{(1)}(\textbf{r})=-\delta{\cal F}_{ex}[\rho]/\delta\rho(\mathbf{r})$ are functions of $r$. Applying an iteration scheme that minimizes $\Omega[\rho]$ we obtain the corresponding density profile $\rho(r)$ for hard spheres at a spherical test particle
using a standard Picard-iteration scheme $\rho^{(i)}(r)=(1-\alpha)\rho^{(i-1)}(r)+\alpha \tilde{\rho}^{(i)}(r)$, where $\alpha\in(0,0.1]$ is a mixing parameter and $\tilde{\rho}^{(i)}(r)=\rho_b\exp\left(-\beta V_{ext}(r)+c^{(1)}(r)+\beta\mu_{ex}\right)$.
 
 We tested the accuracy of our numerics by examining the Gibbs adsorption Eq.\eqref{adsorption-sum-rule} and contact sum rules Eq.\eqref{contact-sum-rule}. Employing $50\times\sigma$ grid points with a resolution of 100 points per radius, we satisfy Eq.\eqref{adsorption-sum-rule} to better than $0.01\%$ and Eq.\eqref{contact-sum-rule} to better than $0.0084\%$ at packing fraction $\eta=0.3$.
 
 From the equilibrium density profile $\rho_{eq}(r)$ and grand potential $\Omega[\rho_{eq}]=\Omega$ we obtain $\mu_{ex}^{\text{DFT}}$ and $\chi_T^{\text{DFT}}$.
 We use Eq.\eqref{relative-devs} to compute the relative deviations $\delta_\mu$ and $\delta_\chi$.
 This procedure is carried out for parameters $A$ and $B$ in a given region and for packing fractions $\eta$ from $0$ to $0.45$.

Having calculated the relative deviations we implement an optimization program that will furnish the optimal choice of parameters. The idea is to search for points $(A,B)$ that deliver small values for $\delta_\mu$ and $\delta_\chi$ over the whole range of  $\eta$ that we consider.

By averaging results for a selection of  (large) values of $\eta$ we select a point that possesses the smallest relative deviation with respect to $\mu_{ex}$ and $\chi_T$. This was achieved by considering the quantity $M(\delta_\mu,\delta_\chi)=\frac{1}{2}(|\delta_\mu|+|\delta_\chi|)$ where we take the absolute values in order to avoid any cancellations arising from differences in sign.

To summarize: we calculate $\delta_\mu(\eta;A,B)$ and $\delta_\chi(\eta;A,B)$ for each choice of parameters $(A, B)$ and three choices of packing fraction $\eta=0.35,\,0.40$ and $0.45$. Then, we obtain the average value $\overline{\delta}_{\mu,\chi}(A,B) = \frac{1}{3}\sum_{\eta=0.35,0.40,0.45}|\delta_{\mu,\chi}(\eta;A,B)|$.
We then find the minimum $\underset{A,\,B}{\text{min}}\,M(\overline{\delta}_\mu, \overline{\delta}_\chi)$, thereby obtaining the optimal point $(A, B)$.

\section{Results}\label{results}
\begin{figure}
\includegraphics[width=0.49\linewidth]{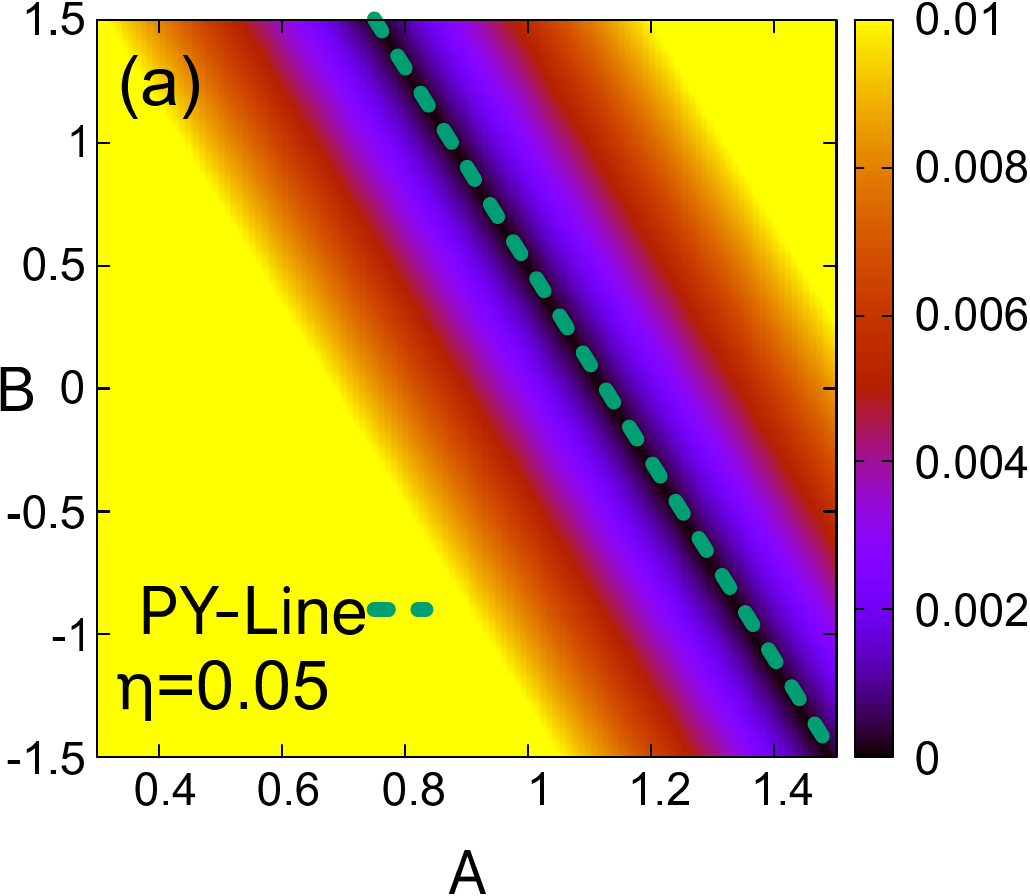}
\includegraphics[width=0.49\linewidth]{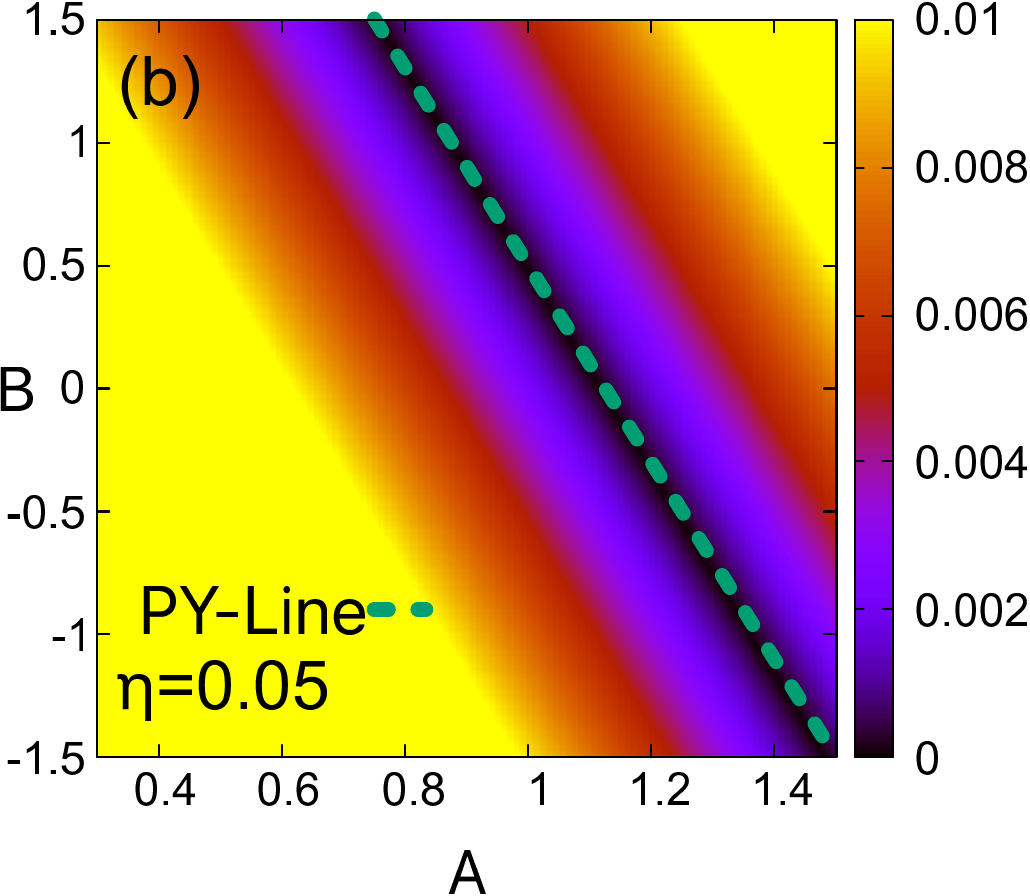}
\caption{\label{fig:muEx_xt_0.05} Relative deviations (a) $|\delta_\mu|$ and (b) $|\delta_\chi|$, defined in Eq.\eqref{relative-devs}, in the parameter plane $(A, B)$ for packing fraction $\eta=0.05$. The color indicates the magnitude of the deviation. The PY-line is given by $8A+2B-9=0$.}
\end{figure}
\begin{figure}
\includegraphics[width=0.49\linewidth]{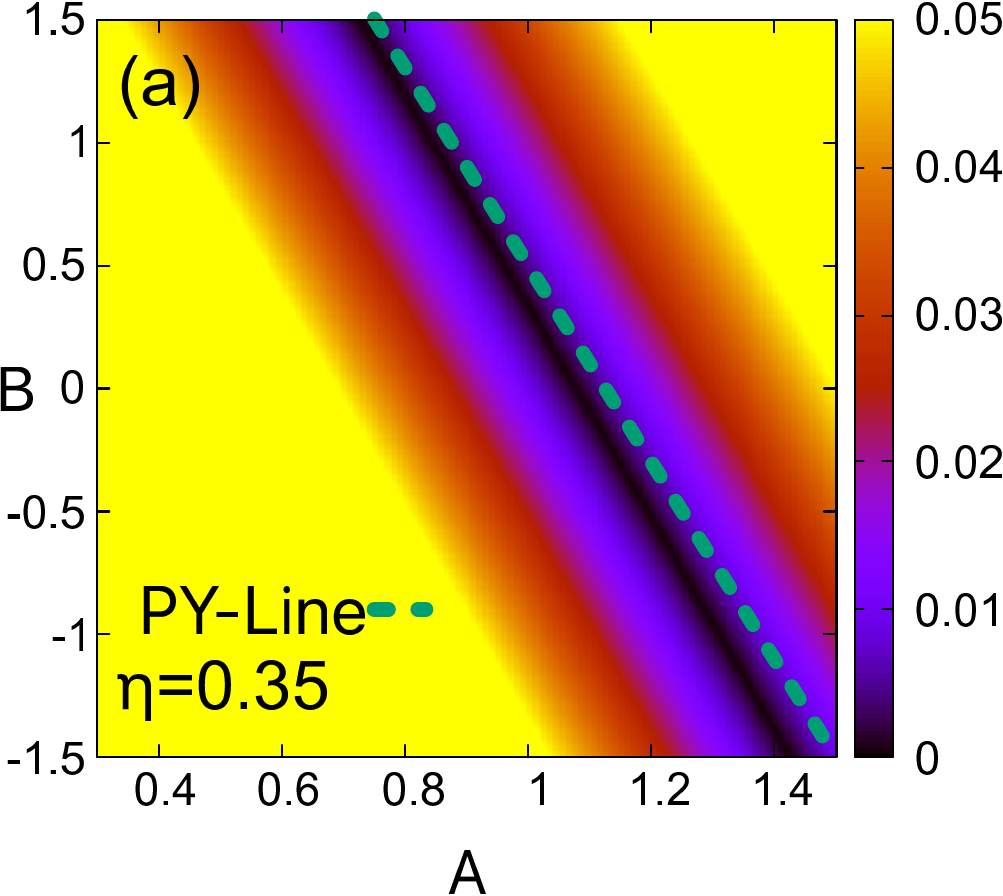}
\includegraphics[width=0.49\linewidth]{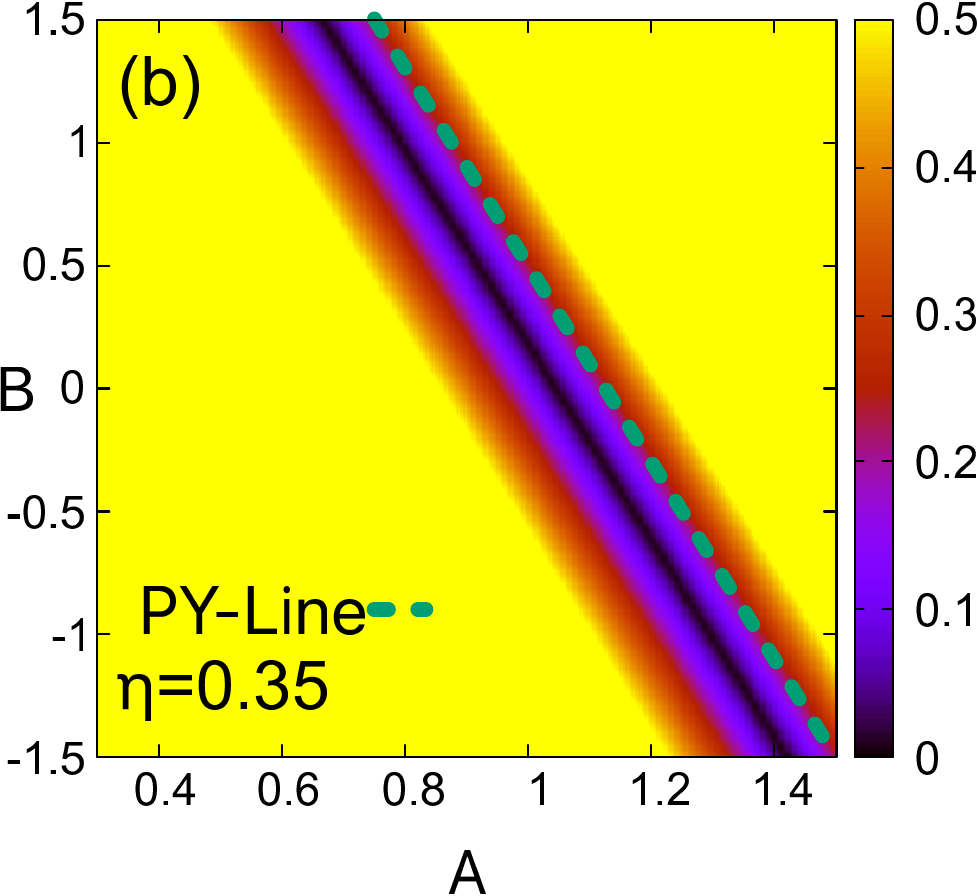}
\caption{\label{fig:muEx_xt_0.35} Relative deviations (a) $|\delta_\mu|$ and (b) $|\delta_\chi|$, defined in Eq.\eqref{relative-devs}, in the parameter plane $(A, B)$ for packing fraction $\eta=0.35$.  The color indicates the magnitude of the deviation.The PY-line is given by $8A+2B-9=0$}
\end{figure}
For each choice of $A$ and $B$ in Lutsko's functional, we compute $\mu_{ex}^{\text{DFT}}$ and $\chi_T^{\text{DFT}}$. It is illuminating to represent the corresponding relative deviations $\delta_\mu$ and $\delta_\chi$, defined in Eq.\eqref{relative-devs}, in a heatmap, Fig.\ref{fig:muEx_xt_0.05} and Fig.\ref{fig:muEx_xt_0.35}. At each point $(A, B)$ in the parameter plane $|\delta_\mu|$ and $|\delta_\chi|$ are displayed in a color code. We add the PY-line, in bright green, given by $8A+2B-9=0$ for reference. We choose to display a region $A\in[0.3, 1.5]$ and $B\in[-1.5, 1.5]$ for our results since i) small deviations occur in the vicinity of the PY-line and ii) the optimal point falls in this range.

In Fig.\ref{fig:muEx_xt_0.05} the packing fraction is very small: $\eta=0.05$ and the relative deviations for both quantities are small. The regions corresponding to the smallest deviations are aligned with the PY-line. This is not surprising as PY is exact to second order in the density. For higher $\eta$ , however, the lines of minimum deviation start to shift  from the PY-line, as can be seen in Fig.\ref{fig:muEx_xt_0.35}, where $\eta=0.35$.

Moreover, we observe that the region of small deviations for $\delta_\chi$ is narrower than for $\delta_\mu$. Note the absolute scales; deviations in the compressibility, $\delta_\chi$, are generally very much larger than those in the excess chemical potential, $\delta_\mu$. We can infer that there is a region with $A\in(1.2,1.4)$ and $B\in(-1.2,-0.9)$ where both relative deviations must be small. 

That we have guaranteed alignment with the PY-line in the low-density regime implies that we do not have to consider low densities/packing fractions in our approach for optimization.

Thus, employing the prescription described in Sec.\ref{NumMeth}, we find the following optimal point 
\begin{equation}\label{optimal-point}
	A = 1.3,\quad B = -1.0,
\end{equation}
which neither lies on the PY-line nor belongs to esFMT$(A, B)$. Lutsko \cite{esMFT-Lutsko} has shown that demanding explicit stability, i.e. $A,B\geq 0$, creates differences from PY in the pair direct correlation function $c^{(2)}(r)$ already at first order in the density, which is unfavorable since PY is exact up to second order. Lutsko proposed a simple choice $A=1$ and $B=0$ which still produces a small error at second order in $\eta$ but improves the virial expansion at higher order \cite{esMFT-Lutsko}.

We present results for this choice as well as for the optimal point Eq.\eqref{optimal-point} and compare with RF and WB.
\begin{figure}[h]
\includegraphics[width=0.49\linewidth]{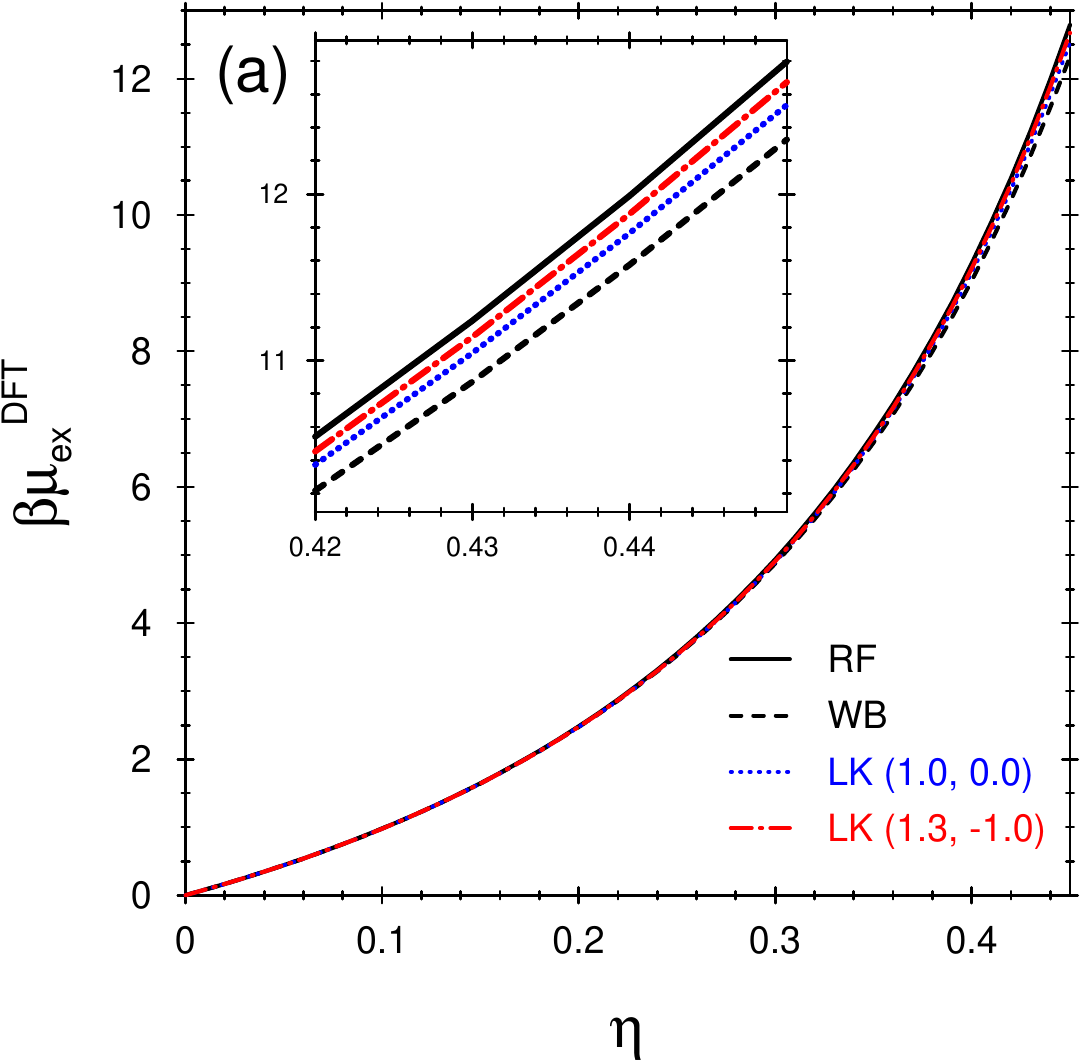}
\includegraphics[width=0.49\linewidth]{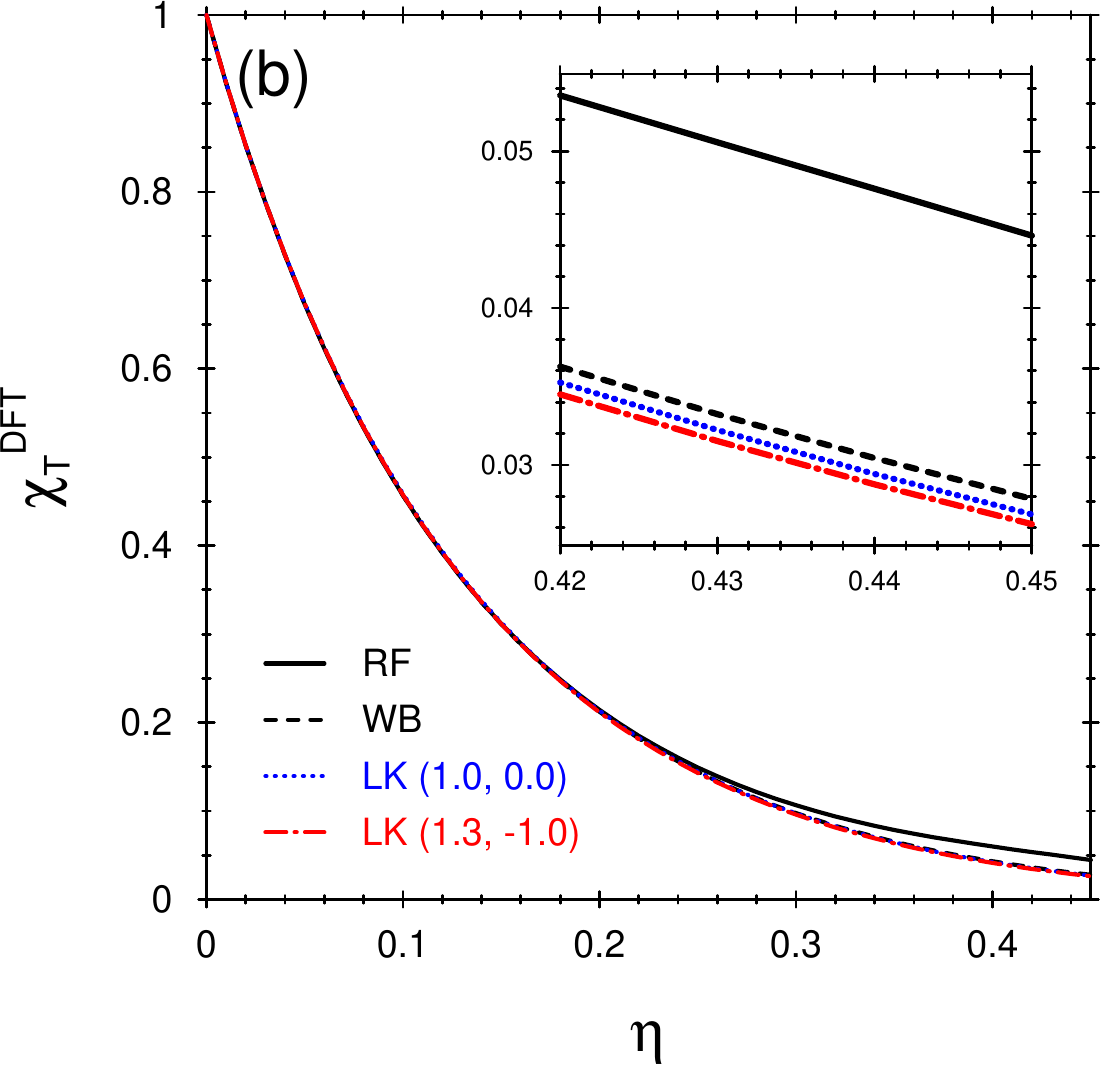}
\caption{\label{fig:sumrule_mu_xt} Density functional results from test-particle calculations: (a) $\beta\mu_{ex}^{\text{DFT}}$ and (b) $\chi_T^{\text{DFT}}$ vs. the packing fraction $\eta$; the insets show the results on an expanded scale for high densities. Four different functionals are considered.}
\end{figure}
In Fig.\ref{fig:sumrule_mu_xt}(a) we see that for $\mu_{ex}^{\text{DFT}}$ the results from the two Lutsko functionals (LK) lie between the RF and WB results for large $\eta$. LK$(1.0, 0.0)$ is closer to WB. The WB results for $\chi_T^{\text{DFT}}$, Fig.\ref{fig:sumrule_mu_xt}(b), lie well below RF and both LK results are below WB, with LK$(1.0,0.0)$ closer to WB.
This turns out to be significant when we examine the overall consistency of the underlying functional in Fig.\ref{fig:sumrule_devs_mu_xt}. Here we plot the relative deviations $\delta_\mu$ and $\delta_\chi$ versus packing fraction. In Fig.\ref{fig:sumrule_devs_mu_xt}(a) we see that results for $\delta_\mu$ for our optimal point LK$(1.3,-1.0)$ are generally more consistent than for LK$(1.0,0.0)$ over the full range of $\eta$. As expected, RF performs better than the LK functionals for $\eta\leq 0.2$. WB exhibits the smallest deviations across the range of $\eta$ . Overall the relative deviations for $\delta_\mu$ are small for the various functionals  even for high densities , typically $<1\%$.

In Fig.\ref{fig:sumrule_devs_mu_xt}(b), where results for $\delta_\chi$ are plotted, we first observe that the magnitude of the deviation is generally much larger than for $\delta_\mu$; note the vertical scales. WB again exhibits small deviations but these grow as $\eta$ approaches 0.45. LK$(1.0,0.0)$ exhibits a large variation with $\eta$ whereas LK$(1.3,-1.0)$ shows a smoother, monotonic variation and $\delta_\chi$ $<0.04$ at $\eta=0.45$. RF performs poorly with $\delta_\chi>0.13$ for $\eta=0.3$ and growing for higher values of $\eta$.

\begin{figure}[h]
\includegraphics[width=0.49\linewidth]{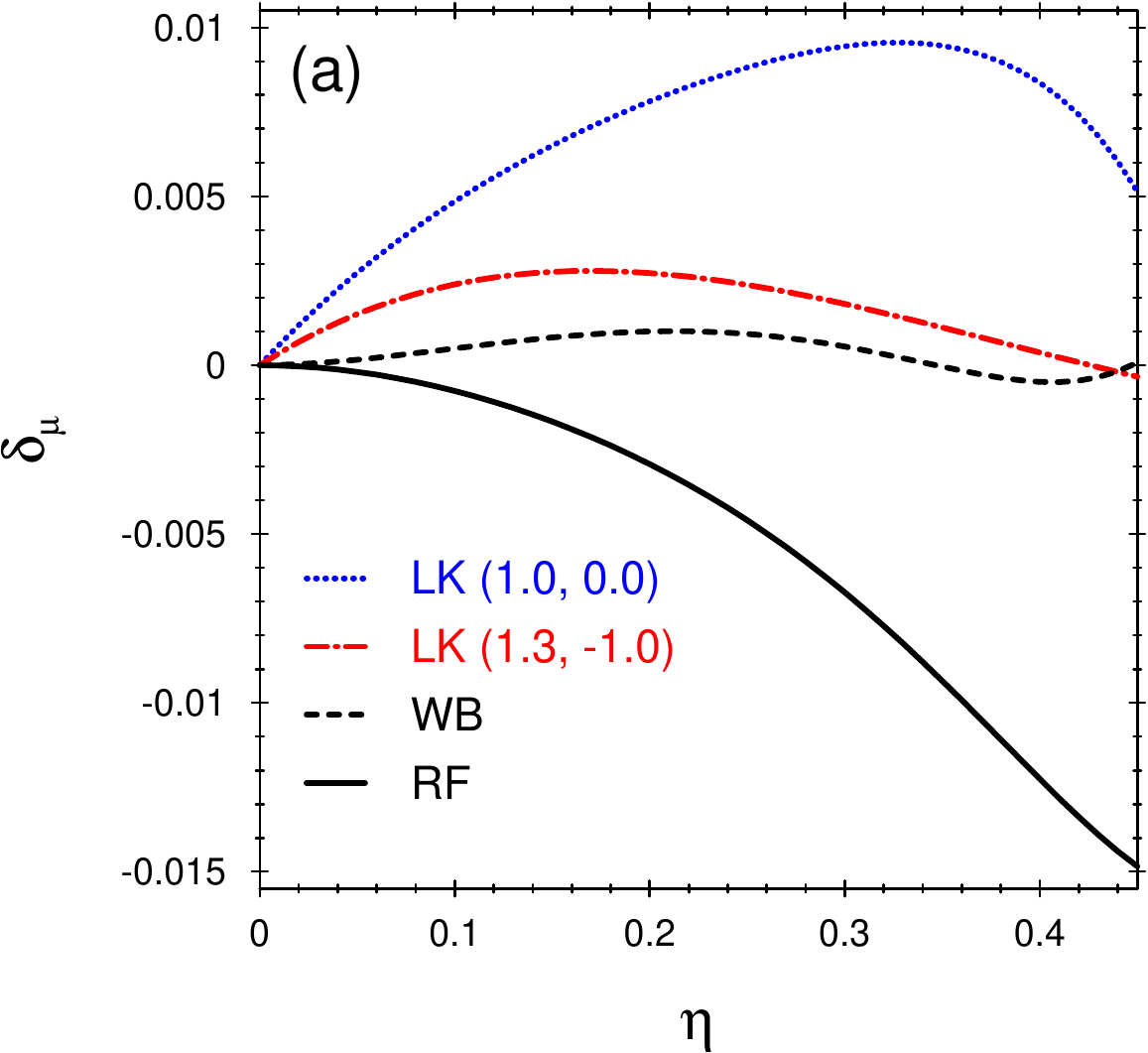}
\includegraphics[width=0.49\linewidth]{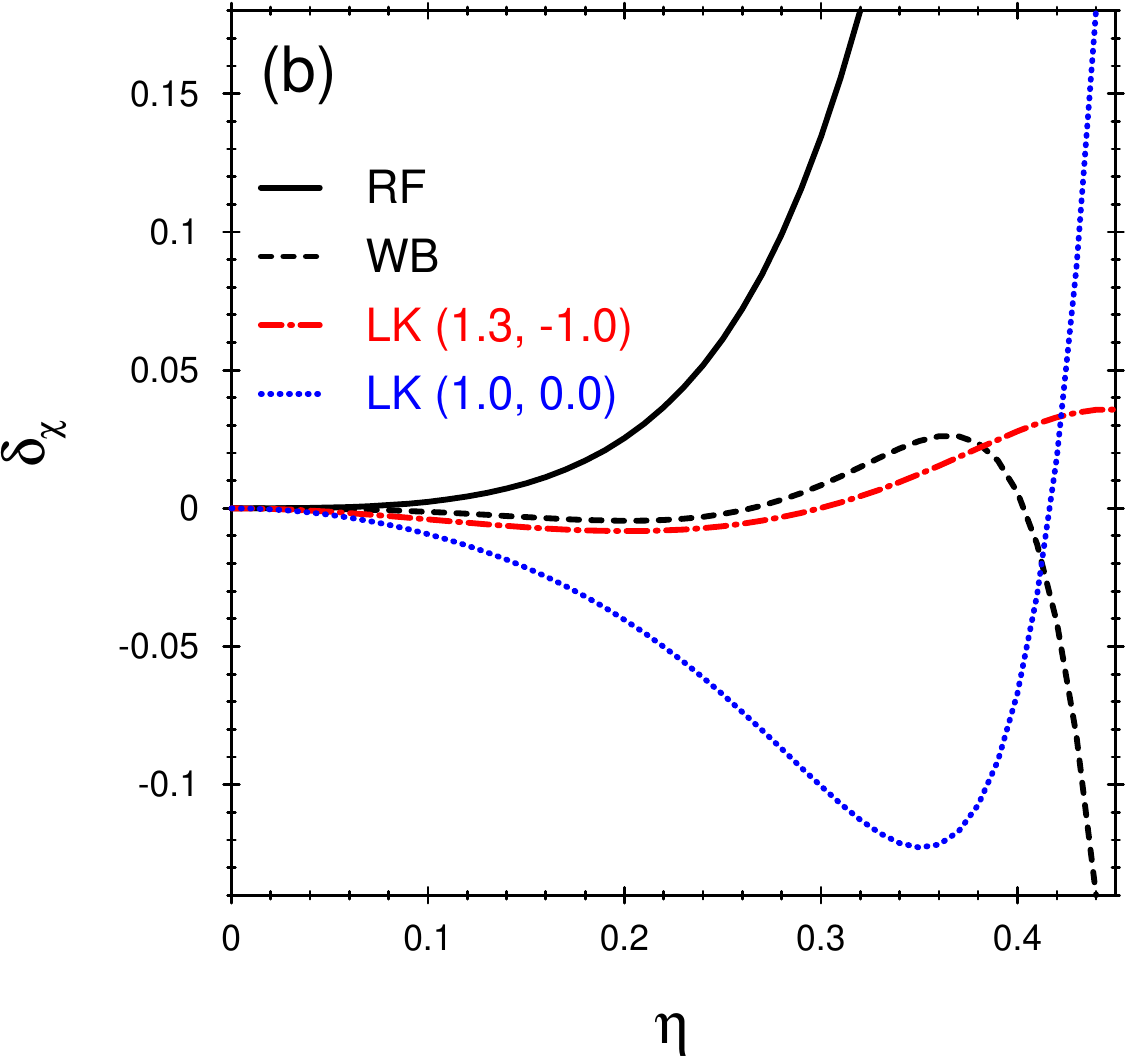}
\caption{\label{fig:sumrule_devs_mu_xt} Relative deviations (a) $\delta_\mu$ and (b) $\delta_\chi$, defined in Eq.\eqref{relative-devs}, vs.the packing fraction $\eta$. Note that for the Rosenfeld functional (RF) and the (original) White Bear version of FMT (WB) the deviations are fixed from the underlying theory. By contrast for the Lutsko functional deviations depend on the parameters $A$ and $B$ and hence can be tuned.We display results for certain choices.}
\end{figure}
Note that from the thermodynamic result, Eq.\eqref{xt-bulk}, we would expect $\chi_T$ for RF to lie below LK(1.3,-1.0), LK(1.0,0.0) and WB. Why this is not the case in the DFT results of Fig.\ref{fig:sumrule_mu_xt}(b) now becomes clear.

Although we have derived our optimal point using large $\eta$ data the relative deviations remain small for lower densities. This is simply a consequence of the fact that the optimal point is close to the PY-line which represents the correct behaviour at small $\eta$. We also note that LK functionals with $8A+2B\neq 9$ create errors of second order in the density as signalled in Fig.\ref{fig:sumrule_devs_mu_xt}(a) where the relative deviation $\delta_\mu$ starts to increase rapidly as $\eta$ increases from zero.

\begin{figure}[h]
\includegraphics[width=0.49\linewidth]{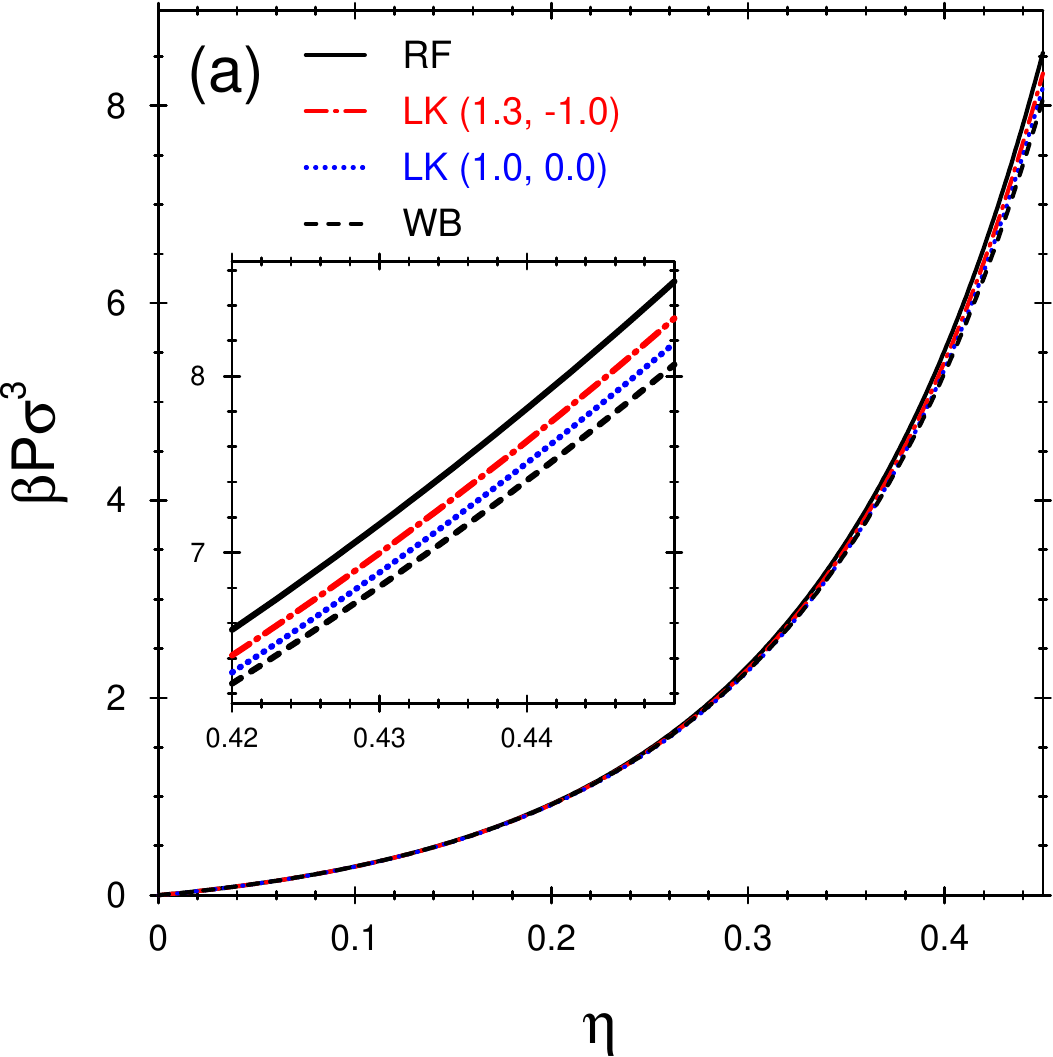}
\includegraphics[width=0.49\linewidth]{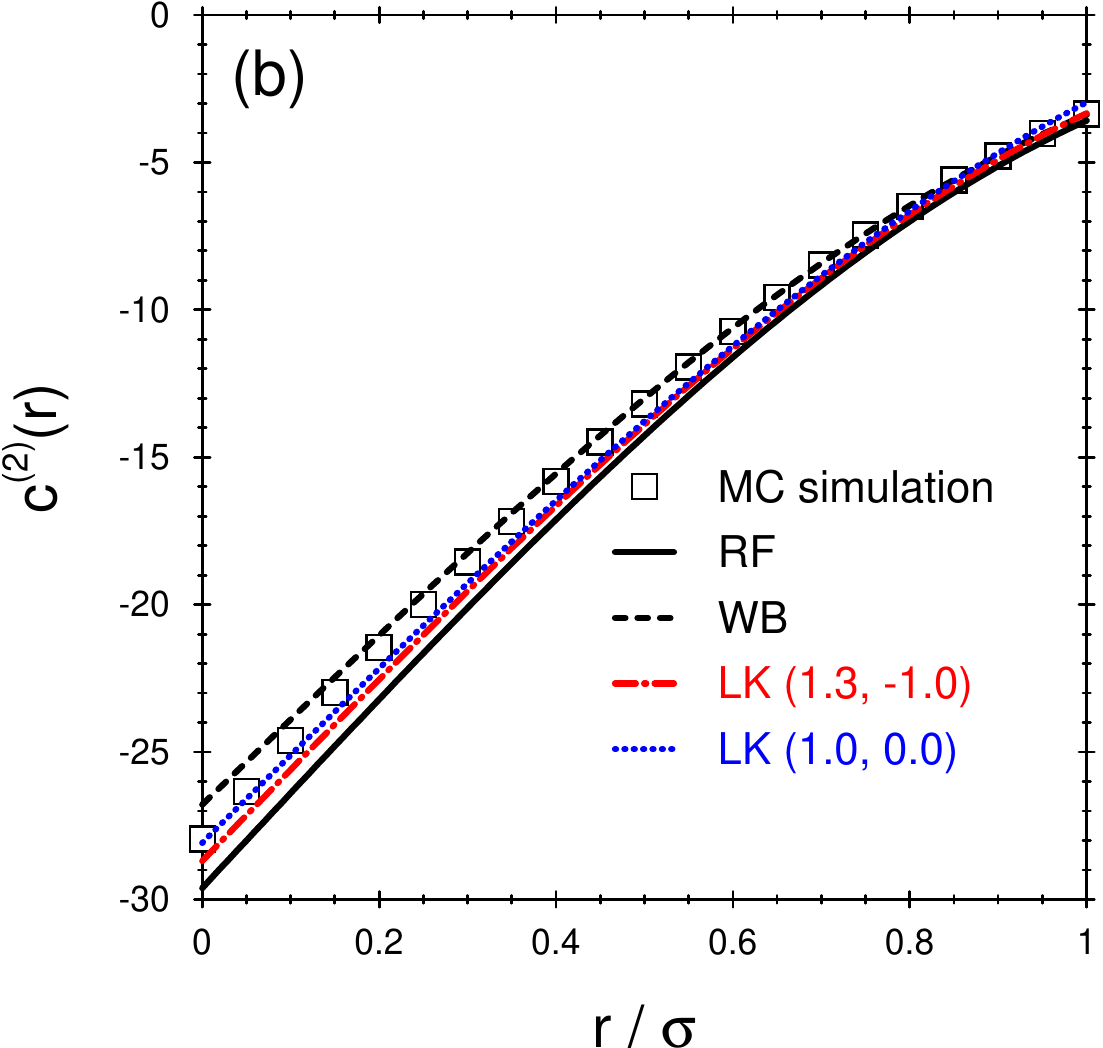}
\caption{\label{fig:pressure} (a) Reduced pressure $\beta P\sigma^3$ vs. the packing fraction $\eta$; the inset shows the results on an expanded scale for high densities. Recall that the WB results are equivalent to those of CS which are very close to simulation. (b) Pair direct correlation function $c^{(2)}(r)$ as a function of $r/\sigma$ for $\eta=0.4189$. DFT results are compared with MC simulation data \cite{Corr-Func-hard-spheres}. Recall $\sigma=2R$ is the HS diameter.In all these theories $c^{(2)}(r)=0$, for $r>\sigma$.} 
\end{figure}

Fig.\ref{fig:pressure}(a) shows the pressure as a function of the packing fraction $\eta$. At high densities the deviations between the different functionals become evident. Both LK$(1.3, -1.0)$ and LK$(1.0, 0.0)$ satisfy $8A+2B<9$. Thus, according to Eq.\eqref{LK-EoS}, the corresponding pressure will be lower than PY. LK$(1.0, 0.0)$ is closer to WB which corresponds to the Carnahan-Starling (CS) EoS, known to be very close to that from computer simulations \cite{Hansen-McDonald}. Adjusting $A$ and $B$ such that the CS result is obtained does not work since the latter corrects the PY result, Eq.\eqref{PY-EoS}, by subtracting a term proportional to $\eta^3$ whereas the additonal term stemming from Lutsko's functional, Eq.\eqref{LK-EoS}, is proportional to  $\eta^2$. However, provided $8A+2B<9$ the corresponding correction lowers the PY pressure thereby coming closer to the accurate CS result. 

In this context it is instructive to examine the virial coefficients. $B_n^{\text{esFMT}}$ from the Lutsko functional, given in Table I of \cite{esMFT-Lutsko}, deviate from the PY virial coefficients by an amount proportional to $C=\frac{1}{3}(8A+2B-9)$ for $n>2$. While PY is exact for $n=2$ and $n=3$, it overestimates the higher order coefficients $B_n^{\text{exact}}$. Since we have $C<0$ for the Lutsko functionals considered here, the corresponding virial coefficients improve upon those of PY. For LK$(1.3,-1.0)$, with $C= -0.2$, $B_4$ is closer to the exact result than that from  LK$(1.0,0.0)$, with $C= -0.33333$. For larger $n$, however, LK$(1.0,0.0)$ yields results that are closer, in keeping with observations above regarding the equation of state. Recall that $B_3 =10+C$ is not exact since $C<0$.

In Fig.\ref{fig:pressure}(b), we compare the pair direct correlation function $c^{(2)}(r)$ between the various functionals and with simulation \cite{Corr-Func-hard-spheres} for $\eta=0.4189$. An explicit formula for $c^{(2)}(r)$ obtained from the Lutsko functional is given in Eq.(24) of Lutsko \cite{esMFT-Lutsko}.(Note, however, that the sign between the PY direct correlation function and the additional term should be negative). Over the whole range of $r/\sigma$ WB exhibits the best agreement with MC simulation whereas LK$(1.0, 0.0)$ performs better than WB for small values of $r/\sigma$.
However, for $r$ very close to $\sigma$ we observe that LK$(1.3, -1.0)$ fits the data from MC simulation better than LK$(1.0, 0.0)$, coinciding with WB. Focusing on $c^{(2)}(r)$ suggests a further strategy for optimization of parameters $A$ and $B$. Choosing $\eta=0.4189$, as in Fig.\ref{fig:pressure}(b), and seeking  the minimal quadratic error between $c^{(2)}(r)$ and simulation data we obtain $A=1.44$ and $B=-1.90$, values quite different from what we found employing the test-particle sum rules. We can also determine $A$ and $B$ by demanding that the values of $c^{(2)}(r)$ at $r=0$ and $r=\sigma$ are fixed by those of the simulation data, yielding $A=1.32$ and $B=-1.33$, i.e. values closer to our optimal point from the sum rules. We remark that LK functionals that are not members of esFMT class seem to fulfill the various optimizations best.
\begin{figure}
\includegraphics[width=0.49\linewidth]{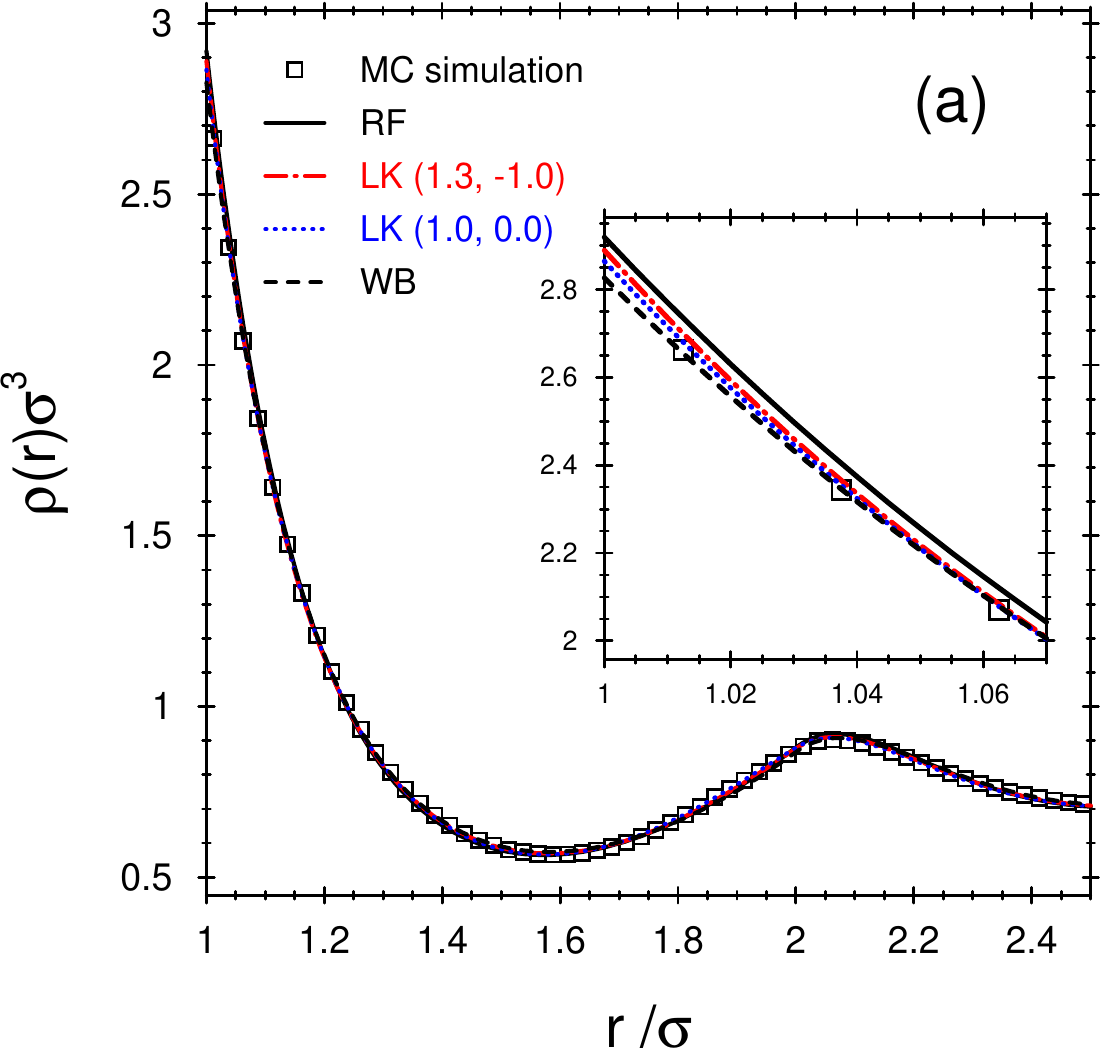}
\includegraphics[width=0.49\linewidth]{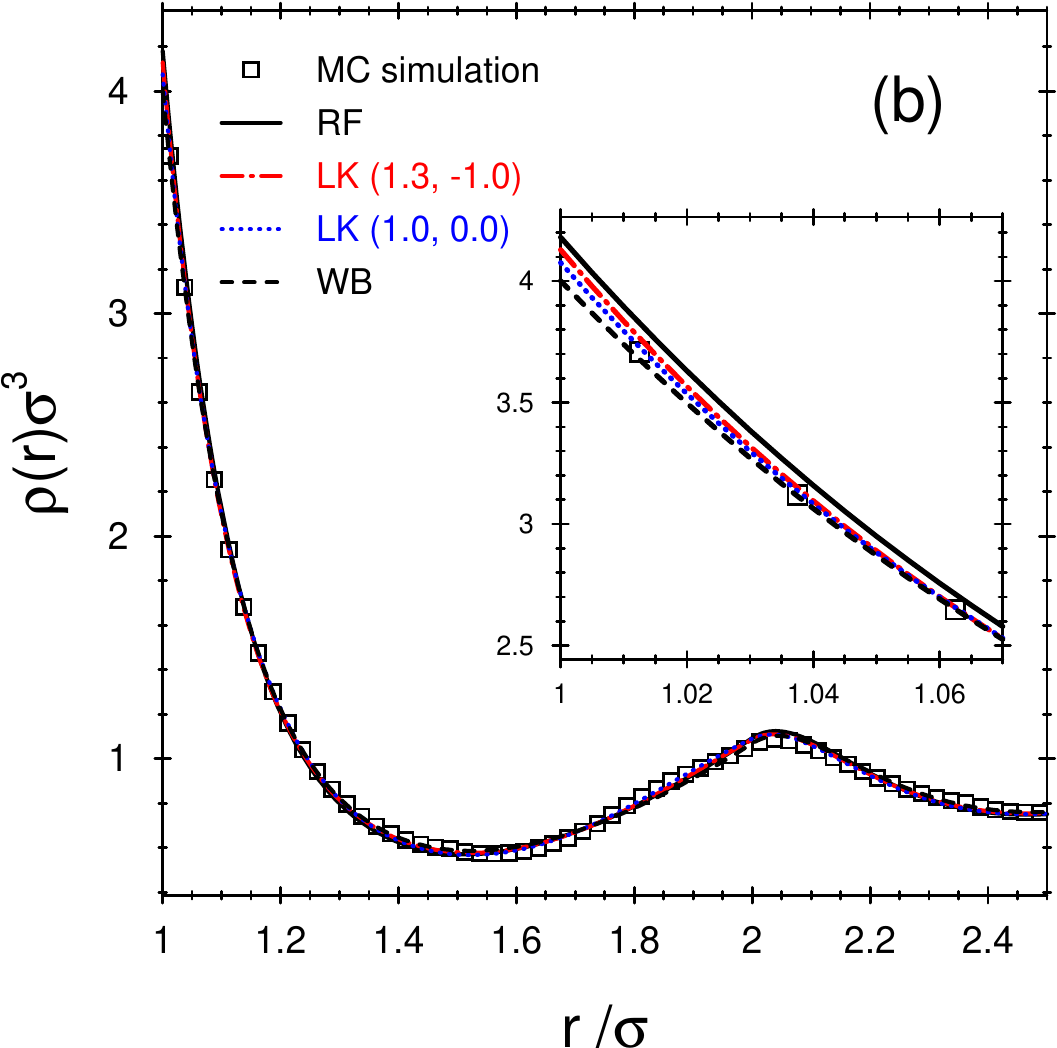}
\caption{\label{fig:spherical-wall} Density profile $\rho(r)$ at a spherical test particle of diameter $\sigma$ for (bulk) packing fractions (a) $\eta=0.40$ and (b) $\eta=0.45$; the inset shows results close to contact on an expanded scale.}
\end{figure}
\begin{figure}
\includegraphics[width=0.49\linewidth]{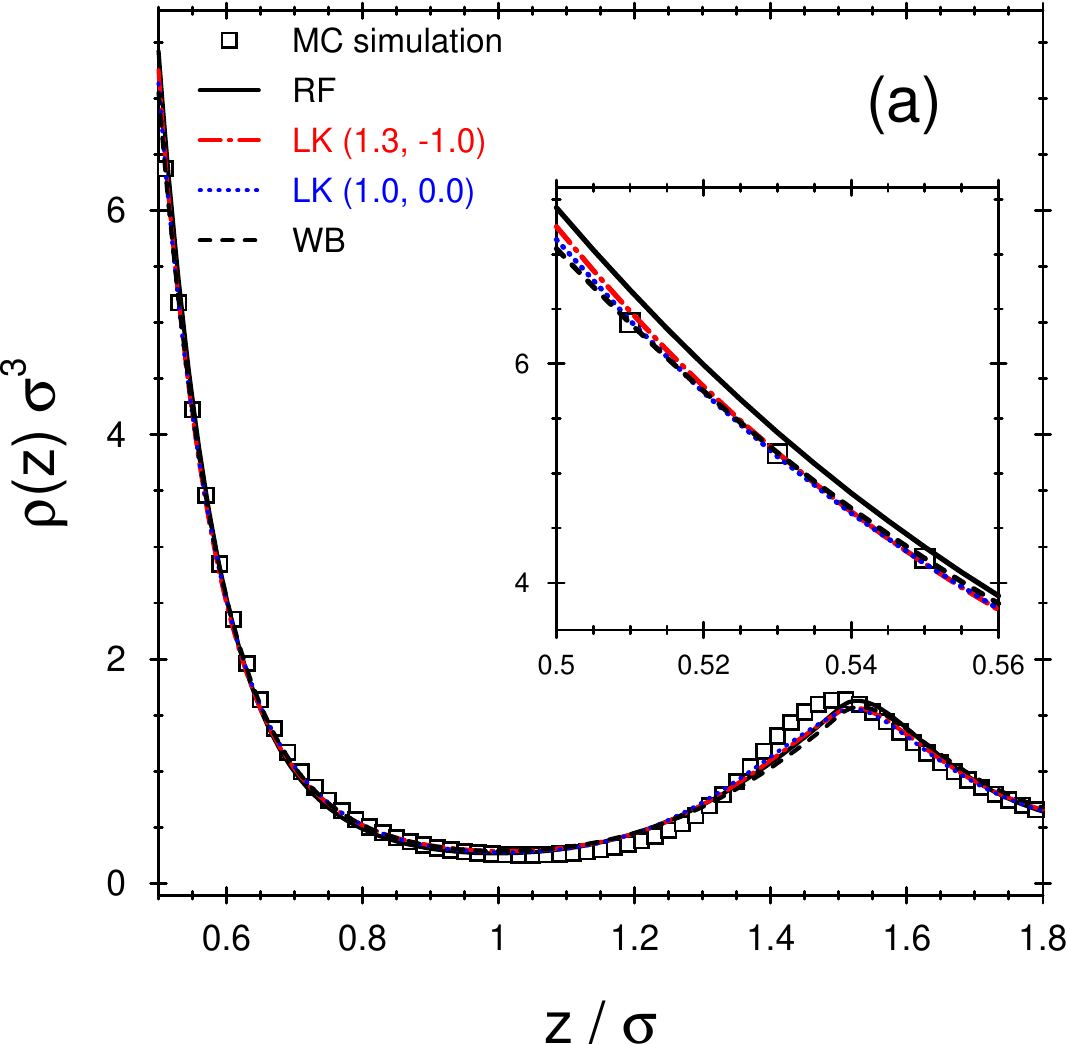}
\includegraphics[width=0.49\linewidth]{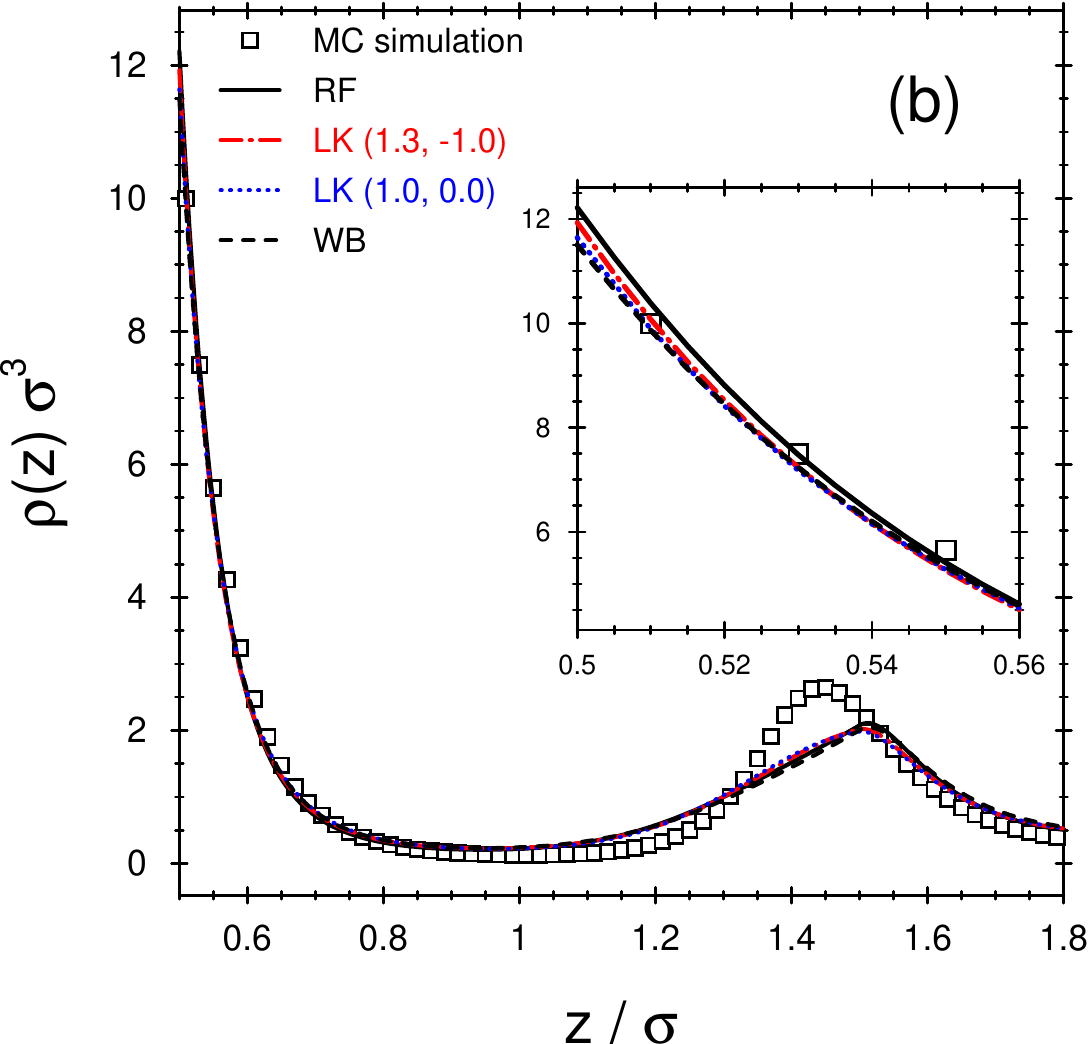}
\caption{\label{fig:hard-wall} Density profile $\rho(z)$ at a planar hard wall for (bulk) packing fractions (a) $\eta=0.4341$ and (b) $\eta=0.4911$; the inset shows results close to the wall on an expanded scale.
The MC data are taken from \cite{Davidchack15}.} 
\end{figure}

Finally, we examine the overall accuracy of the various functionals we have considered by comparing results, with MC simulation, for density profiles $\rho(\mathbf{r})$, for several high values of $\eta$, at a hard spherical test particle \ref{fig:spherical-wall}) Fig.7 and a hard planar wall (Fig.\ref{fig:hard-wall}).Note that the density profile at the spherical test particle is plotted with contact at $r=\sigma$ , as in Fig.1, whereas for the planar hard wall contact is at $z=0.5\times\sigma$.

From Fig.\ref{fig:spherical-wall} we observe that for the four functionals we consider the agreement with our own simulation data is good with slight differences near the first maximum of the density profile, i.e. $r/\sigma\approx 2$. Only in the vicinity of contact with the test particle do deviations become relevant. The density profile from the RF-functional lies above the MC data points whereas results corresponding to LK-functionals stay between RF and WB. As expected, WB captures best the MC data points near the test particle. LK$(1.0, 0.0)$ and LK$(1.3, -1.0)$ both perform better than RF. Of course, this reflects the accuracy of the underlying EoS and the contact value theorem Eq.\eqref{contact-sum-rule}.

For the planar hard wall, Fig.\ref{fig:hard-wall}, we see that deviations from MC are pronounced at the first maximum of the density profile near $z/\sigma\approx 1.5$, especially for $\eta=0.4911$, see Fig.\ref{fig:hard-wall}(b). Deviations also occur close to the wall. Again WB fits the MC data points best,reflecting the fact that it incorporates the CS EOS. It is important to note that there is no reason that LK$(1.3,-1.0)$ should provide an accurate description at the planar wall. Recall this was derived in the test-particle geometry. Fig.\ref{fig:hard-wall} clearly shows that the corresponding planar profile is very close to those from other functionals; differences occur only close to the wall.

\section{Summary and Outlook} \label{sec:conclusion}

Current FMT functionals have served as a powerful tool to  describe hard-sphere mixtures. Building upon the Rosenfeld (RF), the subsequent White Bear (WB) versions of FMT make accurate predictions, compared to computer simulations, for the structure and thermodynamics of such systems. The search for even more accurate functionals continues.With this aim in mind, we have investigated the role of statistical mechanical sum rules. The well-known Gibbs-adsorption theorem, Eq.~(\ref{adsorption-sum-rule}), and the contact theorem, Eq.~(\ref{contact-sum-rule}) are respected by FMT ; they can only provide a test for the numerical accuracy of the DFT minimisation.In this paper we have introduced sum rules for the excess chemical potential $\mu_{ex}$, Eq.~(\ref{excess-grand-pot}), and isothermal compressibility $\chi_T$, Eq.(\ref{iso-compress}), in test-particle geometry, making use of ideas from Percus. These sum rules are not respected by FMT, which implies that they can be used to test the internal consistency of the theory. We employed these sum rules in order to establish the {\em relative deviations} $\delta_\mu$ and $\delta_\chi$, Eq.(\ref{relative-devs}), between quantities calculated from different routes. Results for the RF and WB versions of FMT are shown in Fig.~\ref{fig:sumrule_devs_mu_xt} as black full (RF) and dashed (WB) lines, respectively.We see that WB performs rather well apart from the compressibility at high packing fractions.

Since these two sum rules are not guaranteed to be satisfied by FMT they provide a route to determine the free parameters $A$ and $B$ entering the Lutsko functional, Eq.~(\ref{eq:phi3_LK}) \cite{esMFT-Lutsko}. In this case the relative deviations $\delta_\mu$ and $\delta_\chi$ are functions of the (reservoir) packing fraction $\eta$ and of the  parameters $A$ and $B$. We sought to minimise the deviations. While it would be possible to emphasise one sum rule over the other, we weighted them equally by defining the target of our minimisation (in a machine learning setting this would be called the loss function) as the arithmetic mean of the relative deviations, namely  $M(\delta_\mu,\delta_\chi)=\frac{1}{2}(|\delta_\mu|+|\delta_\chi|)$, which is, of course, still a function of $\eta$, $A$ and $B$. We found that minimising $M$ w.r.t. $A$ and $B$ for a fixed value of $\eta$ results in an improved consistency close to the chosen value of $\eta$, but an increased inconsistency for values of the packing fraction sufficiently far from it. Thus, we chose to average the value of $M$ over three different values of $\eta\in\{0.35,0.40,0.45\}$ and minimised the errors w.r.t. $A$ and $B$; see Sec.~\ref{NumMeth}. The optimal choice of parameters, for the fluid states we consider, are those given in Eq.~(\ref{optimal-point}): $A=1.3$ and $B=-1.0$, which we denote by  LK$(1.3, -1.0)$.

We noted that these parameters differ from those selected by Lutsko \cite{esMFT-Lutsko}, LK$(1.0, 0.0)$, which were determined with the constraint of explicit stability ($A\geq 0$ and $B\geq 0$) and with crystalline phases in mind. Clearly the negative value $B=-1.0$ found by us violates one of the explicit stability conditions. However, this does not necessarily imply that the Lutsko functional, Eq.~(\ref{Lutsko-Functional}), with the parameters $A=1.3$ and $B=-1.0$ is unstable. 

Our optimal choice of parameters for the fluid state, LK$(1.3, -1.0)$, do not fall on the Percus-Yevick (PY) line, defined by the relation $8A+2B=9$ where the equation of state, Eq.~(\ref{LK-EoS}), reduces to the PY compressibility equation of state which slightly overestimates the pressure of a hard-sphere fluid. For LK$(1.3, -1.0)$ the third term in Eq.~(\ref{LK-EoS}) is negative so that the pressure is closer to that given by the accurate Carnahan-Starling (CS) equation of state. Fig.\ref{fig:pressure}(a) shows that the parameters suggested by Lutsko, LK$(1.0, 0.0)$, result in an equation of state even closer to CS. Similar observations can be made when comparing pair direct correlation functions $c^{(2)}(r)$ with simulation; see Fig.\ref{fig:pressure}(b).

Overall we find that the performance of the Lutsko functional with parameters $A=1.3$ and $B=-1.0$ is promising. Regarding consistency, we find that the relative deviation for the chemical potential, plotted in Fig.~(\ref{fig:sumrule_devs_mu_xt})(a), is significantly smaller than for RF and the Lutsko functional with LK$(1.0, 0.0)$, and is comparable with those of the WB version of FMT.  The relative deviation for the isothermal compressibility, plotted in Fig.~(\ref{fig:sumrule_devs_mu_xt})(b), is not only smaller than those of the RF and the Lutsko functional with LK$(1.0, 0.0)$, but is also smaller than that of the WB version of FMT, especially at higher values of $\eta$.

As mentioned earlier, the Lutsko FMT functional Eq.\eqref{eq:phi3_LK} is an extension of the tensor version \cite{Tarazona-Rosenfeld-1997} of the Rosenfeld functional \cite{Free-energy-model-of-hard-spheres}. It is important to recognize that there are several ways to modify and potentially improve the structure of FMT functionals. The tensor version of the Rosenfeld functional \cite{Tarazona-Rosenfeld-1997} and the Lutsko functional Eq.\eqref{eq:phi3_LK} focused on the dependence of $\Phi_3$ on the scalar, vectorial and tensorial weighted densities $n_2$, $\mathbf{n}_2$, and $\mathbf{T}$, respectively, in order to improve the stability of the crystalline phase. However,it is also possible to change the dependence of $\Phi_3$ (and possibly of $\Phi_2$) on the scalar weighted density $n_3$, in order to improve the underlying equation of state \cite{Tarazona02,FMT-Revisited-and-WB,Structures-of-hard-sphere-fluids-from-mod.-FMT}. Following the point of view presented in Ref.~\cite{FMT-for-hard-sphere-mixtures} of the {\em FMT toolbox}, the modifications suggested by Lutsko \cite{esMFT-Lutsko} can easily be applied to other versions of FMT, such as the White-Bear version \cite{FMT-Revisited-and-WB,Structures-of-hard-sphere-fluids-from-mod.-FMT}  or White-Bear mark II \cite{WB-mark-II}. Since these versions of FMT are somewhat more consistent (see results for the unmodified original White-Bear version in Fig.~\ref{fig:sumrule_devs_mu_xt}) one can expect an even higher degree of self-consistency than we found here for the Lutsko functional.

Looking ahead, we should recall that the sum rules we have introduced apply for \textit{any} pair potential. We chose to focus on one-component hard spheres since it is for this case that we have the most accurate excess free energy functionals. More generally, a long standing problem in classical DFT is how to incorporate accurately attractive interactions between particles. Most applications employ a simple mean-field treatment, that ignores correlations, e.g. Ref.~\cite{DFT-of-Non-uniform-fluids}. A recent theory \cite{Tschopp20}, based upon a Barker-Henderson treatment of the bulk\cite{Barker76}, improves upon this by including correlations in the hard-sphere reference system. It would be interesting to examine the degree to which such theories satisfy the sum rules, employing the measures we introduced here. We might also enquire about a recent machine learning (ML) approach that learns the one-body direct correlation functional from training with inhomogeneous one-body density profiles, for hard-spheres \cite{Sanmueller23} and for the Lennard-Jones fluid \cite{Sanmueller24}.This approach then determines the bulk pair structure via functional differentiation and the bulk free energy via functional line integration.  How well does this approach , and that of other ML for DFT e.g.\cite{Dijkman24}, satisfy our sum rules?

\begin{acknowledgments}
We thank Mary Coe, Josh Robinson and Jim Lutsko for fruitful discussions. R.E. acknowledges support of the Leverhulme Trust, grant no. EM-2020-029/4.
\end{acknowledgments}

\bibliographystyle{apsrev}
\bibliography{lutsko.bib}

\end{document}